\documentclass[prl,twocolumn,preprintnumbers,reprint,amsmath,amssymb]{revtex4-1}

\usepackage{epsfig,amsfonts}
\usepackage{color}

\def\be{\begin{equation}}
\def\ee{\end{equation}}
\newcommand{\comment}[1]{}

\def\bea{\begin{eqnarray}}
\def\eea{\end{eqnarray}}

\def\ben{\begin{enumerate}}
\def\een{\end{enumerate}}

\def\bea{\begin{eqnarray}}
\def\eea{\end{eqnarray}}

\begin{document}

\title{Integrability-based analysis of the hyperfine-interaction-induced\\ decoherence in quantum dots}
\author{Alexandre Faribault and Dirk Schuricht }
\affiliation{Institut f\"{u}r Theorie der Statistischen Physik, RWTH Aachen University and JARA - Fundamentals of Future Information Technology, 52056 Aachen, Germany}

\date{\today}
\begin{abstract}
Using the algebraic Bethe ansatz in conjunction with a simple Monte Carlosampling technique, we study the problem of the decoherence of a central spin coupled to a nuclear spin bath. We describe in detail the full crossover from strong to weak external magnetic field field, a limit where a large non-decaying coherence factor is found. This feature is explained by Bose-Einstein-condensate (BEC)-like physics which also allows us to argue that the corresponding zero frequency peak would not be broadened by statistical or ensemble averaging. 
\end{abstract}
\maketitle

Using the spin of a single electron (or hole) trapped in semiconductor-based quantum dots has been a long-standing proposal for a possible implementation of a qubit~\cite{qbit} for which single-spin readout and coherent control~\cite{exp} is now possible. In these setups, the isotropic Fermi contact hyperfine interaction of the trapped central spin with the bath of nuclear spins present in the substrate is know to be the essential source of decoherence leading, over time, to a loss of the information encoded in any prepared state of the central spin. Denoting by $\vec{S}_0$ the central spin-$\frac1{2}$, by $\vec{I}_j$ the $N$ nuclear spins, and by $g$ and $g_n$ the respective couplings to the external magnetic field $h$, the Hamiltonian reads
\bea
H = g h S^z_0 + g_\text{n} h\sum_{j=1}^{N} I_j^z + \sum_{j=1}^N A_j \vec{S}_0\cdot\vec{I}_j.
\label{csH}
\eea
Dropping an irrelevant constant the first two terms can be simplified~\cite{coish2004,bortz2010_2} to $BS_0^z$ with $B=(g-g_\text{n})h$.

In the regime of strong magnetic field $B > A$, with $A$ being $N$ times the largest $A_j$, the time evolution of the central spin $\langle\vec{S}_0(t)\rangle$ can be described perturbatively~\cite{coish2004,khaetskii2003,PT,cywinski2010,coish2010} in the flip-flop terms $\sum_jA_j(S_0^+I_j^-+\text{h.c.})$. In a detailed study to fourth order Coish, Fischer, and Loss~\cite{coish2010} found that the Larmor precession undergoes a typical exponential decay which is supplemented by additional low-frequency envelope modulations. In the opposite limit of weak (or intermediate) fields a variety of methods including semiclassical approaches, exact diagonalization, or the algebraic Bethe ansatz(ABA) have been applied~\cite{khaetskii2003,cywinski2010,nonPT1,erlingsson2004,tdmf1,nonPT2}. For unpolarized spin baths at $B=0$, extrapolating a semi-classical approach~\cite{erlingsson2004} to the continuum limit, as well as time-dependent mean field theory~\cite{tdmf1} for very large systems with $N=16\ 000$ showed logarithmic decay $\langle S_0^z(t)\rangle\sim 1/\ln(t)$ at long times. However, all approaches in the weak field limit are either based on mean-field or semiclassical methods or are restricted to either very small system sizes $N\le 20$, specific bath polarizations, or the short-time behavior.  

In this work we aim to overcome these limitations by developing a hybrid method based on the ABA in combination with a direct Monte Carlo (MC) sampling of the exact eigenstates of the system, which we apply to the transverse relaxation of the central spin. The method provides a nonperturbative full quantum treatment giving us access to the complete crossover from strong to weak field without any restriction on the initial polarization of the bath. Because of the trivial time evolution in the eigenbasis, it also provides access to real-time dynamics at arbitrarily long times  without accumulated errors. While it is only usable for modest system sizes, recent developments have drastically sped up the computation time. In this work we treat systems containing up to $N=48$ nuclear spins whose Hilbert space is $\sim 2\times10^8$ times larger than the ones previously treated with exact methods~\cite{cywinski2010,nonPT1}.

For the dynamics of $\langle S^+(t)\rangle$ we find the following conclusions: (i) At strong to intermediate fields we observe exponential decay with weak modulations and thus confirm the perturbative picture~\cite{coish2010}. (ii) For smaller fields, these weak modulations become dominant leading to long-lived slow oscillations. (iii) As $B \to 0$, an initial rapid decay is followed by the formation of a nondecaying coherent fraction whose amplitude and phase is in one-to-one correspondence with the initial state of the central spin. This lack of decay is explained as a consequence of Bose-Einstein condensate (BEC)-like physics. (iv) We argue that the nondecaying fraction survives ensemble averaging and is at most logarithmically dependent on finite size effects. Consequently, we expect it to be experimentally observable.

To be specific, we will consider nuclear spins $\frac1{2}$ although this restriction is not imposed by integrability or the numerical method. The couplings are chosen as $A_j = \frac{A}{N} e^{-(j-1)/N}$ corresponding to a Gaussian electronic wave function in a 2D quantum dot \cite{coish2004}. We study in detail the time evolution of an initial coherent superposition of the central spin: $\frac{1}{\sqrt{2}}\left(\left|\Uparrow\right> + \left|\Downarrow\right>\right)$ related to the $T_2$ transverse relaxation time. We assume the initial condition to be a product state $\left|\Psi_0\right> =  \frac{1}{\sqrt{2}}\left(\left|\Uparrow\right> + \left|\Downarrow\right>\right) \otimes \left|\Psi_\text{bath}\right>$, with $\left|\Psi_\text{bath}\right>$ describing the nuclear spins. Narrowing techniques \cite{narrowing} will typically allow one to create a superposition of (nearly-)degenerate eigenstates of the Overhauser operator $h^z=\sum_{j=1}^NA_j I^z_j$. Provided there is no special phase relation between them, the density matrix can be reduced to its diagonal terms, leading to an averaged sum over the represented $h^z$ eigenstates. Here we avoid the average by using a single specific zero-polarization "maximal entropy" state: $\left|\Psi_\text{bath}\right> = \left|\downarrow_1 \ \uparrow_2 \ \downarrow_3 \ \uparrow_4 \cdots \right>$ with the eigenvalue $h^z_\text{init} = \sum_{j=1}^N(-1)^jA_j/2$. Having the up-pointing spins uniformly spread out over the full range of available coupling strengths should make this state sufficiently generic to capture the dominant features of a diagonal ensemble average. We will confirm the validity of this particular choice by explicitly comparing randomly generated initial nuclear configurations. 

Our main quantity of interest is the central spin coherence factor which, projecting on the eigenbasis of $H$,  can be written as 
\bea
\langle S_0^+(t)\rangle=\left<\Psi(t)\right|S_0^+\left|\Psi(t)\right> = \sum_{m,n} \frac{C_{m,n}}{2} e^{i (\omega_m-\omega_n)t}, 
\label{ooft}
\eea
where $C_{m,n} =
 \left<\Uparrow;\Psi_\text{bath} \right. \left|\psi_m\right> \left<\psi_m\right|S_0^+\left|\psi_n\right> \left<\psi_n\right.  \left|\Downarrow;\Psi_\text{bath}\right>$ with $\left|\psi_{m}\right>$ and $\left|\psi_{n}\right>$ denoting eigenstates of the Hamiltonian (\ref{csH}) with energies $\omega_m$ and $\omega_n$ respectively. Since $S_0^z+\sum_{j=1}^N I^z_j$ is conserved, $\left|\psi_{m}\right>$ must contain one more up-spin than $\left|\psi_{n}\right>$. 
 
Any eigenstate of the central spin model (\ref{csH}) is entirely defined by $M$ complex rapidities $\{\lambda_1 ... \lambda_M\}$ which need to be a solution of a system of $M$ coupled non-linear algebraic equations: the Bethe equations \cite{bortz2010_2}. Defining $\epsilon_k=-1/A_k$ and $\epsilon_0 =0$, the corresponding (unnormalized) eigenstate is obtained by the repeated action, for each rapidity, of a generalized creation operator $\mathrm{S}^+(\lambda_i) \equiv \frac{S_0^+}{\lambda_i-\epsilon_0}+\sum_{k=1}^N\frac{I_k^+}{\lambda_i-\epsilon_k}$, i.e.
$\left|\left\{\lambda_1 ... \lambda_M\right\}\right> = \prod_{i=1}^M \mathrm{S}^+(\lambda_i) \left|\Downarrow; \downarrow \downarrow ... \downarrow\right>$.
The corresponding  eigenenergy is then given by
\bea
\omega(\left\{\lambda_1 ... \lambda_M\right\}) =  \frac{1}{2}\sum_{i=1}^M \frac{1}{\lambda_i} - \frac{B}{2} - \frac{1}{4}\sum_{j=1}^N\frac{1}{\epsilon_j}.
\label{ener}
\eea
Any eigenstate can therefore be pictured as containing a given number $M$ of individual quasiparticles.  Each of them is described by one single complex parameter $\lambda_i$, which specifies both the spin profile and its energy $\mathrm{Re}\left[1/\lambda_i\right]$.

\begin{figure}
\includegraphics[angle=90,width=\columnwidth]{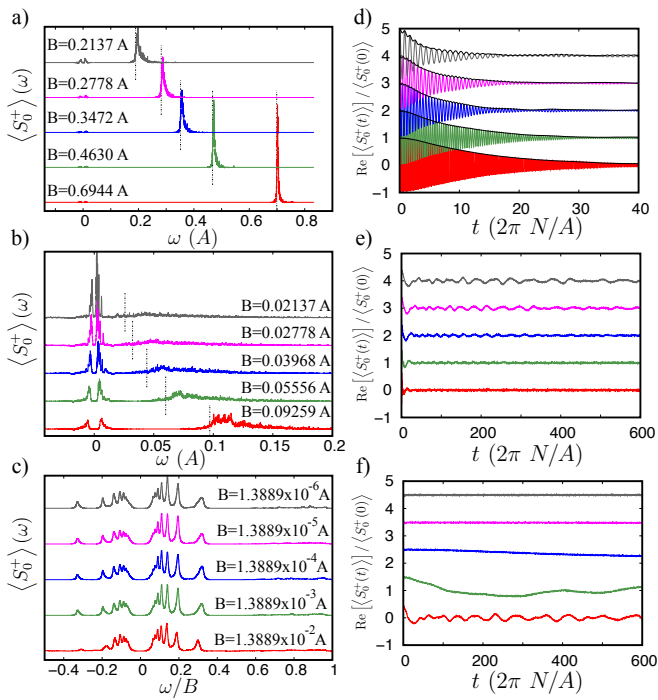}
\caption{Left: Spectrum $\left<S^+_0(\omega)\right>$ in the crossover and weak magnetic field regimes. The dashed lines mark the "bare" Larmor frequencies $B+h^z_\text{init}$. Panel c) is plotted in terms of the rescaled frequency $\omega/B$ and shows only the low-frequency structure. Right: The corresponding real-time evolution of the coherence factor $\left<S^+_0(t)\right>$. Each curve has an offset of 1 compared to the previous one. Black lines in panel (d) are the norms $\left|\left<S^+_0(t)\right>\right|$ which, in the perturbative regime, correspond to the envelope function computed in \cite{coish2010}.}
\label{resplots}
\end{figure}

In this work, instead of finding rapidities themselves we solve for a different set of variables $\Lambda(\epsilon_i) =\sum_{j=1}^M \frac{1}{\epsilon_i-\lambda_j}$, which can be shown to obey a simple set of quadratic equations~\cite{babelon,gritsev1}. Any given solution to these equations is found starting from the trivial $B = \infty$ solutions where an ensemble of $M$ spins are pointing up and the remaining $N-M$ are pointing down. These configurations are deformed by a stepwise ramping of the $1/B$ parameter until the desired $B$ value is reached~\cite{gritsev1} . Finally, it was recently shown that the scalar products matrix elements defining $C_{m,n}$ in Eq. (\ref{ooft}) can be written, in terms of $\Lambda(\epsilon_i)$, as determinants of $N+1\times N+1$ matrices~\cite{determinant}. The resulting fast algorithm allows us to define a probability $P_{m,n} \equiv \left| C_{m,n}\right|$ for any pair of eigenstates $(m,n)$ and perform the double sum in Eq. (\ref{ooft}) using a simple Metropolis algorithm (see \cite{buccheri} for another example of combining MC with ABA). Starting from a randomly selected pair of $B=\infty$ eigenstates, we first deform them to the desired finite-$B$ eigenstates $\left|\psi_m\right>,\left|\psi_n\right>$ and compute the probability $P_{m,n}$, frequencies $\omega_m,\omega_n$ and the sign $s_{m,n}=\mathrm{sgn}(C_{m,n})$. A new pair is then generated by randomly selecting one of the two $B=\infty$ configurations and minimally changing it by exchanging a randomly selected pair of up and down spin. Deforming this new configuration to finite $B$ we compute $P_{m,n'}$ (assuming state $n$ was modified) and accept the new pair $(m,n')$ with probability $\mathrm{min}\left(1, \frac{P_{m,n'}}{P_{m,n}}\right)$. Repeating the procedure generates a list of $\Omega$ configurations $(m_\alpha,n_\alpha)$ distributed according to $P_{m_\alpha,n_\alpha}$ such that 
\bea
 \frac{\left<S^+_0(t)\right>}{\left<S^+_0(0)\right>}= \lim_{\Omega \to \infty}  \frac{\sum_{\alpha=1}^{\Omega} s_{m_\alpha,n_\alpha}e^{i(\omega_{m_\alpha}-\omega_{n_\alpha}) t} }{\sum_{\alpha=1}^{\Omega} s_{m_\alpha,n_\alpha} },
\eea
\noindent which can be normalized by the known initial value of the coherence factor. Figure \ref{resplots} presents the spectrum $\left<S^+_0\right>(\omega)$ and its Fourier transform $\left<S^+_0(t)\right> $ for a wide range of external magnetic fields covering the full nonperturbative crossover $B_\text{fluc} \lesssim B \lesssim A$, where 
$B_\text{fluc}=(\sum_{j=1}^NA_j^2)^{1/2}$ is a typical Overhauser field due to spin fluctuations~\cite{tsyplyatyev2011}, all the way down to very weak magnetic fields. All plots are obtained for an ensemble of $N=36$ nuclear spins by sampling $\Omega=10^7$ configurations. For any finite size system, the spectrums consist of a series of delta peaks which are smoothed into Lorentzians of width $0.001 B$ in the plots. We stress that this broadening is much smaller than the width of the peaks seen on the graphs; i.e., the peaks are  composed of a large number of contributions at similar frequencies.

The spectrum is basically characterized by two structures. First we find a peak around the "bare" Larmor  frequency $B + h^z_\text{init}$ given by the total effective magnetic field felt by the central spin. For the strongest magnetic fields this sharp peak is the dominant feature whose nearly Lorentzian line shape leads to exponentially decaying oscillations. As the magnetic field is lowered its width increases giving rise to faster decoherence. In addition, there is a low-frequency structure which carries a very low weight at strong fields but leads to slow modulations of the envelope function. These modulations were previously observed in the fourth-order perturbative treatment~\cite{coish2010} and are now confirmed by our exact results. 

When lowering the field below $B_\text{fluc}\approx 0.1\,A$ the low-frequency structure becomes the dominant feature, eventually taking a scaling form of $\omega / B$ for sufficiently weak fields [see Fig.~\ref{resplots}.(c)]. The low-frequency structure has a finite width $(\propto B)$ and, as $B \to 0$, collapses into a delta peak. In the long-time dynamics, this leads to the slowly decaying low-frequency oscillations shown in Fig.~\ref{resplots} (e), whose period and lifetime can be made arbitrarily large since the real-time evolution becomes a function of $Bt$. As $B \to 0$ this ultimately gives rise to a nondecaying fraction representing nearly $0.5$ of the initial value. At $B=0$ the total coherence factor $\left<S^+_0\right> + \sum_{j=1}^N \left<I^+_j\right>$ is conserved, hence the lost central spin coherence is transferred to the nuclear spins polarizing them along the central spin's initial orientation and ultimately locking the system into a non-decaying steady state. 

Here, by working with exact eigenstates of the model we get valuable insight in the processes involved which allow us to demonstrate the complete absence of long-time decay, i.e. in contrast to other approaches~\cite{erlingsson2004,tdmf1} we do not observe a $1/ \ln( t )$ decay. In the $B\to 0$ limit, eigenstates contain two independent subsets of quasiparticles. A state-dependent number $0 \le r \le M$ of rapidities diverge as $\lambda_i \approx L_i/B + \mathcal{O}(1)$ ($L_i$ being roots of a Laguerre polynomial \cite{altshuler}). The corresponding excitations become part of a Bose-Einstein-like condensate of quasi-particles. Being created by the same operator $\mathrm{S}^+(\lambda_i \to \infty) \propto  S^+_0+\sum_{k=1}^N I^+_k$, they are indeed all identical and all have zero-energy $\mathrm{Re}\left[1/\lambda_i\right] \propto B  + \mathcal{O}(B^2)$. On top of this condensate one finds $M-r$ additional excitations with finite energies. They are characterized by finite $\lambda_i \approx \lambda_i^0+\mathcal{O}(B)$ whose proximity with some particular values of the inverse couplings $\left\{\epsilon_i\right\}$ leads to a localized excitation profile. Retaining the same set of localized excitations, one can actually form new zero-field eigenstates with the same energy by adding any number of quasiparticles to the condensate. This particular structure of the exact eigenstates is another manifestation of BEC-like physics which echoes the behavior of other integrable Gaudin models such as the Dicke model's superradiance \cite{dicke} or the Richardson model's superconductivity \cite{richardson}. 

\begin{figure}[t]
\includegraphics[angle=90,width=\columnwidth]{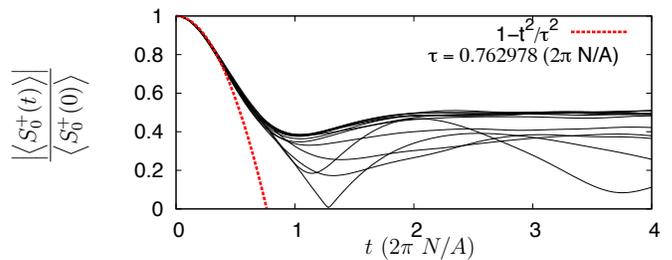}
\caption{Short time dynamics of the norm $\left|\left<S^+_0(t)\right>\right|$ for intermediate to weak fields. The nine values of magnetic fields $B<B_\text{fluc}$ shown in Fig.~\ref{resplots} are plotted with black lines ($B/A \in \left[1.3889\cdot10^{-6},0.05556\right]$). The red line is a quadratic fit with parameter $\tau\sim (A/N)^{-1}$. For both GaAs and Si:P, using values from \cite{coish2004} gives $\tau\approx 1 \mu \text{s}$.}
\label{shorttime}
\end{figure}

In Eq. (\ref{ooft}), any pair of eigenstates which only differ by one additional condensed quasi-particle will contribute to the nondecaying ($\omega=0$) fraction at $B=0$. At weak finite fields they lead instead to a low-frequency ($\omega = \omega_m-\omega_n \propto B$) contribution explaining the scaling seen in panel (c) of Fig. \ref{resplots}. Any other pair of eigenstates gives a finite frequency contribution. These are spread over a large energy band, leading to the initial rapid decay. The eigenenergies are dominated by the finite rapidities (localized excitations), weak magnetic fields only give subleading corrections to this feature. As shown in Fig. \ref{shorttime} for $B\lesssim B_\text{fluc}$, this results in $B$-independent early dynamics well described by a quadratic decay law.
 
Experimental observation of the nonperturbative long-time contributions; e.g., the low-frequency oscillations, requires two conditions. It needs to have a lifetime sufficiently longer than the initial decay and, at the same time, it has to carry a sufficiently large weight. From a rough analysis of the real-time numerical data, fields $B \le 0.03 A$ ($\sim 0.1\,\text{T}$ for GaAs or $\sim 0.0015\,\text{T}$ for Si:P using~ \cite{coish2004} $A/(g^* \mu_B) = 3.5\,\text{T}$ and $A/(g^* \mu_B) = 0.05\,\text{T}$, respectively) would be sufficient to exhibit oscillations with an amplitude of $\sim 10 \%$ of the initial coherence factor and a lifetime long enough compared to the initial rapid decay [see the $B=0.02778 A$ curve in Fig.~\ref{resplots} (e)]. These oscillations would be clearly distinguishable from any perturbative result since their frequency is nearly an order of magnitude lower than the Larmor frequency. Since we are looking for a clear measurable signature, the resulting threshold is at much weaker fields than the naive limit of validity of perturbation theory $B\sim A$.

\begin{figure}[b]
\includegraphics[angle=90,width=\columnwidth]{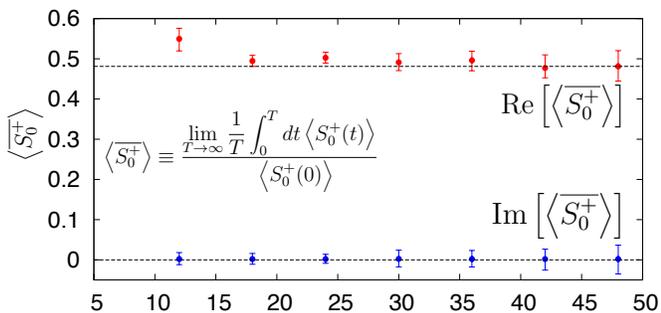}
\caption{The nondecaying coherent fraction as a function of the number of nuclear spins at $B/A = 1.04168\times10^{-6}$. We sample $\Omega=10^7$ configurations. Error bars indicate the magnitude of the fluctuations which are due to a mixture of finite size effects and MC error. Dashed lines are guides to the eye at 0 and 0.4812 (long-time value for $N=48$).}
\label{finitesize}
\end{figure}

In Fig.~\ref{finitesize} we show the long-time averaged contribution for a variety of system sizes. We see that even for modest system sizes ($N \ge 20$) the finite-size discretization does not seem to affect strongly the total weight carried by the low-frequency contributions. The error bars and system sizes treated here do not exclude a slow logarithmic reduction $\sim 1/\ln{(\beta N)}$ of the total fraction as was observed in a finite-size semiclassical treatment~\cite{erlingsson2004} of the longitudinal decay, eventually leading to $~1/\ln(t)$ decay in the continuum limit~\cite{erlingsson2004}. In contrast, our full quantum mechanical treatment includes the discreteness of the spin bath and shows the nondecaying fraction to be observable for finite systems. Indeed, even an assumed $\sim 1/\ln{(\beta N)}$ reduction would yield an observable nondecaying fraction for experimentally relevant system sizes ($N\approx 10^2$ in Si:P or $N\approx 10^5$ in GaAs).

\begin{figure}[t]
    \includegraphics[angle=90,width=\columnwidth]{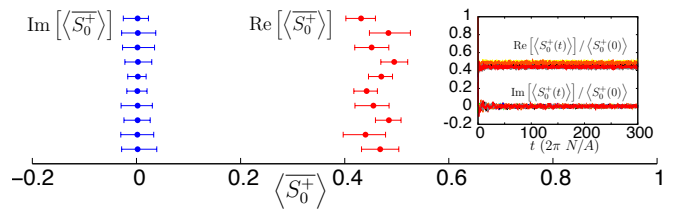}
\caption{The long-time averaged coherent fraction for 10 randomly generated initial nuclear spin configurations. The meaning of the error bars is as in Fig.~\ref{finitesize}. Inset: Corresponding real-time evolution. Magnetic field is $B/A=1.3889\times10^{-6}$ and N=36.}
\label{multistate}
\end{figure}
Considering that even a physically narrowed system should be described in terms of a diagonal ensemble average, in Fig.~\ref{multistate} we compare the weak field dynamics for an ensemble of 10 unpolarized ($M=N/2$) initial bath states obtained by choosing a random set of $M$ up-pointing nuclear spins. This is less restrictive than the eigenstate content of a narrowed state since the initial Overhauser eigenvalue $h^z_\text{init}$ is not fixed. While we observe slight variations in the total non-decaying fraction, it systematically stays above 0.4 and maintains the X-Y plane phase of the initial central spin orientation. Initializing the central spin at a different point on the Bloch sphere $\alpha \left|\Uparrow\right>+\beta \left|\Downarrow\right>$ would only lead to a multiplicative factor $\left<S^+_0(t)\right> = \alpha \beta\left<S^+_0(t)\right>_{(\Uparrow+\Downarrow)}$ which still conserves the relative phase. Albeit averaging may lead to a slightly lowered total fraction, the common phase obtained for all nuclear spin configurations means that diagonal ensemble averages $\left<S^+_0(t)\right> \equiv \sum_{n} \gamma_n \left<\left\{\epsilon\right\}_n\right|S^+_0(t)\left|\left\{\epsilon\right\}_n\right>$, be they performed  for a narrowed or even infinite temperature state, would still yield a large nondecaying fraction. In fact, when projecting any typical initial state onto the weak-field eigenbasis, one will populate a large fraction of the available eigenstates. Since a large majority of them have a finite number of condensed quasiparticles, so will the resulting quantum superposition. On average, one can then systematically expect to find a finite fraction of particles whose dynamics, being frozen by their zero energy, will retain information about the initial state.

 Moreover, the existence of this condensate is completely independent of the set of couplings constants and hence the dot geometry. Therefore, the zero-frequency peak which leads to a nondecaying fraction, should not exhibit any inhomogeneous broadening in experiments which involve an average over different quantum dots. Neither the geometry nor the initial nuclear spin configuration can broaden then delta peak. This is in stark contrast to recent optical spin noise experiments~\cite{osn} where a Lorentzian line shape was observed. However, a direct comparison to our results is not possible considering the experimental work was carried out on hole-spin based samples for which the hyperfine coupling is strongly anisotropic. In the isotropic system considered here a finite lifetime would require the addition of integrability breaking terms to the Hamiltonian \eqref{csH}. The weak dipolar coupling between the nuclear spins could play such a role although on relatively long time scales $\tau_{dd} \approx 10^{-4}$ s in GaAs; another mechanism would be quadrupole couplings to the nuclear spins~\cite{tdmf2}. In addition, while we argued that ensemble fluctuations in the nuclear Overhauser field would not lead to a finite lifetime, it should be noted that local fluctuations in the external magnetic field would induce broadening due to the $\omega/B$ dependency of the spectrum. This demonstrates that, at weak field, the strong correlations between the central and nuclear spins lead to a clear distinction between external magnetic field and the internal Overhauser field.

In conclusion, we have developed a nonperturbative method to study the dynamics in the central spin model which is applicable to arbitrary magnetic fields and initial bath polarizations and provides access to the real-time dynamics at arbitrary long times. We applied this method to the transverse spin relaxation which, at weak fields, shows initial rapid quadratic decay followed by a large arbitrarily long-lived finite coherent fraction resulting from the BEC-like behavior of the system.

We would like to thank  H. Bluhm, J.-S. Caux, L. Fritz, D. Loss, V. Meden, O. Tsyplyatyev, E. Yuzbashyan, and particularly, W. A. Coish for valuable comments and discussions. This work was supported by the German Research Foundation (DFG) through the Emmy-Noether Program under SCHU 2333/2-1 and through SFB631, SFB-TR12, and the Excellence Cluster Nanosystems Initiative Munich (NIM)  while A.F. was affiliated with the Ludwig-Maximilians-Universit\"{a}t, M\"{u}nchen, Germany


\begin{thebibliography}{999}

\bibitem{qbit}
 D. Loss and D. P. DiVincenzo, Phys. Rev. A {\bf 57}, 120 (1998);
 R. Hanson, L. P. Kouwenhoven, J. R. Petta, S. Tarucha, and L. M. K. Vandersypen, Rev. Mod. Phys. {\bf 79}, 1217 (2007).

\bibitem{exp}
J. R. Petta, A. C. Johnson, J. M. Taylor, E. A. Laird, A. Yacoby, M. D. Lukin, C. M. Marcus, M. P. Hanson, and A. C. Gossard, Science {\bf 309}, 2180 (2005);
F. H. L. Koppens, C. Buizert, K. J. Tielrooij, I. T. Vink, K. C. Nowack, T. Meunier, L. P. Kouwenhoven, and L. M. K. Vandersypen, Nature (London) {\bf 442}, 766 (2006);
F. H. L. Koppens, K. C. Nowack, and L. M. K. Vandersypen, Phys. Rev. Lett. {\bf 100}, 236802 (2008);
M. Pioro-Ladri\`{e}re, T. Obata, Y. Tokura, Y. Shin, T. Kubo, K. Yoshida, T. Taniyama, and S. Tarucha, Nature Phys. {\bf 4}, 776 (2008).

\bibitem{coish2004}
W. A. Coish and D. Loss, Phys. Rev. B {\bf 70}, 195340 (2004).

\bibitem{bortz2010_2}
M. Bortz, S. Eggert, and J. Stolze, Phys. Rev. B {\bf 81}, 035315 (2010).

\bibitem{khaetskii2003} A. V. Khaetskii, D. Loss, and L. Glazman, Phys. Rev. Lett. {\bf 88}, 186802 (2002);
Phys. Rev. B {\bf 67}, 195329 (2003).

\bibitem{PT}
C. Deng and X. Hu, Phys. Rev. B {\bf 73}, 241303(R) (2006);
\emph{ibid.} {\bf 78}, 245301 (2008);
W. Yao, R.-B. Liu, and L. J. Sham, \emph{ibid.} {\bf 74}, 195301 (2006);
W. A. Coish, J. Fischer, and D. Loss, \emph{ibid.} {\bf 77}, 125329 (2008);
\L. Cywi\'{n}ski, W. M. Witzel, and S. Das Sarma, \emph{ibid.} {\bf 79}, 245314 (2009);
V. V. Dobrovitski, A. E. Feiguin, R. Hanson, and D. D. Awschalom, Phys. Rev. Lett. {\bf 102}, 237601 (2009).

\bibitem{coish2010}
W. A. Coish, J. Fischer, and D. Loss, Phys. Rev. B {\bf 81}, 165315 (2010).

\bibitem{cywinski2010}
\L. Cywi\'{n}ski, V. V. Dobrovitski, and S. Das Sarma, Phys. Rev. B {\bf 82}, 035315 (2010).

\bibitem{erlingsson2004}
S. I. Erlingsson and Y. V. Nazarov, Phys. Rev. B {\bf 70}, 205327 (2004).

\bibitem{nonPT1}
J. Schliemann, A. V. Khaetskii, and D. Loss, Phys. Rev. B {\bf 66}, 245303 (2002);
V. V. Dobrovitski and H. A. De Raedt, Phys. Rev. E {\bf 67}, 056702 (2003);
M. Bortz and J. Stolze, Phys. Rev. B {\bf 76} 014304 (2007);
M. Bortz, S. Eggert, C. Schneider, R. Stubner, and J. Stolze, \emph{ibid.} {\bf 82}, 161308(R) (2010).

\bibitem{tdmf1}
K. A. Al-Hassanieh, V. V. Dobrovitski, E. Dagotto, and B. N. Harmon, Phys. Rev. Lett. {\bf 97}, 037204 (2006). 

\bibitem{nonPT2}
G. Chen, D. L. Bergman, and L. Balents, Phys. Rev. B {\bf 76}, 045312 (2007);
E. Barnes, \L. Cywi\'{n}ski, and S. Das Sarma,  Phys. Rev. Lett. {\bf 109}, 140403 (2012).

\bibitem{narrowing}
A. Greilich, D. R. Yakovlev, A. Shabaev, Al. L. Efros, I. A. Yugova, R. Oulton, V. Stavarache, D. Reuter, A. Wieck, and M. Bayer, Science {\bf 313}, 341 (2006);
D. J. Reilly, J. M. Taylor, J. R. Petta, C. M. Marcus, M. P. Hanson, and A. C. Gossard, \emph{ibid.} {\bf 321}, 817 (2008);
I. T. Vink, K. C. Nowack, F. H. L. Koppens, J. Danon, Y. V. Nazarov, and L. M. K. Vandersypen, Nature Phys. {\bf 5}, 764 (2009);
X. Xu, W. Yao, B. Sun, D. G. Steel, A. S. Bracker, D. Gammon, and L. J. Sham, Nature  (London) {\bf 459}, 1105 (2009). 

\bibitem{babelon}
O. Babelon and D. Talalaev, J. Stat. Mech. ({\bf2007}) P06013. 

\bibitem{gritsev1}
A. Faribault, O. El Araby, C.Str\"{a}ter, and V.Gritsev, Phys. Rev. B {\bf 83}, 235124 (2011);
O. El Araby, V.Gritsev, and A. Faribault, \emph{ibid.} {\bf 85}, 115130 (2012).

\bibitem{determinant}
A. Faribault and D. Schuricht, J. Phys. A {\bf 45}, 485202 (2012).

\bibitem{buccheri}
F. Buccheri, A. De Luca, and A. Scardicchio, Phys. Rev. B {\bf 84}, 094203 (2011). 

\bibitem{tsyplyatyev2011}
O. Tsyplyatyev and D. Loss, Phys. Rev. Lett. {\bf 106}, 106803 (2011)


\bibitem{altshuler}
E. A. Yuzbashyan, A. A. Baytin, and B. L. Altshuler, Phys. Rev. B {\bf 68}, 214509 (2003). 

\bibitem{dicke} 
R. H. Dicke, Phys. Rev. {\bf 93}, 99 (1954);
K. Hepp and E. H. Lieb, Ann. Phys. {\bf 76}, 360 (1973);
Y. K. Wang and F. T. Hioe, Phys. Rev. A {\bf 7}, 831 (1973).

\bibitem{richardson}
R. W. Richardson, J. Math. Phys. {\bf 6}, 1034 (1965);
J. von Delft and R. Poghossian, Phys. Rev. B {\bf 66}, 134502 (2002);
L. Amico and A. Osterloh, Ann. Phys. (Berlin) {\bf 524}, 133 (2012).

\bibitem{osn}
Y. Li, N. Sinitsyn, D. L. Smith, D. Reuter, A. D. Wieck, D. R. Yakovlev, M. Bayer, and S. A. Crooker, Phys. Rev. Lett. {\bf 108}, 186603 (2012).

\bibitem{tdmf2}
N. A. Sinitsyn, Y. Li, S. A. Crooker, A. Saxena, and D. L. Smith, Phys. Rev. Lett. {\bf 109}, 166605 (2012).

\end{thebibliography}
\end{document}